\pdfoutput=1
\documentclass[3p,times,twocolumn]{elsarticle}

\usepackage{ecrc}

\volume{00}

\firstpage{1}

\journalname{Nuclear and Particle Physics Proceedings}

\runauth{Victor Feuillard}

\jid{nppp}

\jnltitlelogo{Nuclear and Particle Physics Proceedings}

\usepackage{amssymb}

\usepackage[figuresright]{rotating}

\begin{document}

\begin{frontmatter}



\dochead{}

\title{Charmonium production in Pb-Pb collisions measured by ALICE at the LHC}


\author{Victor Feuillard for the ALICE collaboration}

\address{Commissariat \`a l'\'{E}nergie Atomique, IRFU, Saclay, France.}

\begin{abstract}
The ALICE experiment is dedicated to the study of the Quark-Gluon Plasma (QGP), a state of matter where, due to high temperature and density, quarks and gluons are deconfined. One of the probes studied to investigate this state of matter is the production of charmonium states, such as the J/$\psi$ and the $\psi$(2S). Indeed, the presence of the QGP is expected to modify the charmonium production yields, due to a balance between the color screening of the charm quark potential and a recombination mechanism. A suppression of the production yields in heavy ion collisions with respect to pp collisions scaled by the mean number of binary nucleon-nucleon collisions  was observed by ALICE in Pb-Pb collisions at a centre-of-mass energy per nucleon pair $\sqrt s\textsubscript{NN} = 2.76$ TeV. The observed suppression is smaller than the one found by PHENIX at an order of magnitude lower collision energy. This behavior can be explained by a stronger contribution from recombination processes at LHC than at lower energies. In this presentation, we report on new results for the charmonium production in Pb-Pb collisions measured at forward rapidity with the the muon spectrometer of ALICE at $\sqrt s\textsubscript{NN} =5.02$ TeV and their comparison with previous results and model predictions.
\end{abstract}

\begin{keyword}
QGP \sep heavy ion \sep quarkonium \sep J/$\psi$ \sep muon spectrometer \sep ALICE \sep LHC

\end{keyword}

\end{frontmatter}


\section{Introduction}
\label{}
The Quark-Gluon Plasma is a state of matter predicted by lattice QCD where quark and gluons are deconfined. Recent theoretical studies show that the transition temperature to the QGP at a net baryonic chemical potential $\mu_B=0$ is around 155 MeV \cite{Bazavov:2011nk}. There is a particular interest in studying the QGP since models predict that it was the state of matter at the early stages of the Universe (around $\tau=1$ $\mu$s). Experimentally it is possible to create a QGP using ultra-relativistic heavy ion collisions, like at RHIC or the LHC, but only within a short period of time ($\sim$10 fm/\textit{c} at LHC) and a very small volume ($\sim$10\textsuperscript{4} fm\textsuperscript{3} at LHC) \cite{Aamodt:2011mr}. Due to this short time and volume, it is impossible to observe the QGP directly and we measure indirect probes.
Charmonium resonances, as J/$\psi$ and the $\psi$(2S), which are bound states of a $\rm{c\bar{c}}$ pair, are among the most direct signatures for the QGP formation. Theory predicts that charmonia are suppressed in a QGP due to the colour screening: because of the presence of free colour charges in the medium, the $\rm{c\bar{c}}$ pair cannot easily bind \cite{Matsui:1986dk}. What is also interesting is that the difference between binding energies leads to a sequential melting of the different charmonium states as a function of temperature. When reaching the transition temperature, the $\psi$(2S), which has the lowest binding energy, will be suppressed first, whereas J/$\psi$ will be then suppressed at a higher temperature.
However, a competing mechanism can also occur, namely recombination: if there are enough $\rm{c\bar{c}}$ pairs produced, and they thermalize in the QGP, then quarkonia can be regenerated by the recombination of these $\rm{c\bar{c}}$ quark pairs, resulting in an increase of the charmonium yields \cite{BraunMunzinger:2000px}. 

\section{The ALICE Experiment}
The ALICE detector is composed of two main parts, the central barrel and the muon arm \cite{Aamodt:2008zz} \cite{Abelev:2014ffa}. ALICE measures charmonium in both the central barrel and in the muon spectrometer, in the dielectron and in the dimuon decay channel, respectively. As the measurements discussed in this document were made with the muon arm, we will focus on it. The muon arm has a rapidity coverage of $-4.0<\eta<-2.5$ and is composed of 5 stations of tracking chambers, 2 stations of trigger chambers, a dipole magnet and several absorbers, one in the front to reject the background, mainly hadrons; one in front of the trigger chambers used as muon filter, and one around the beam pipe. The Inner Tracking System (ITS) is used to determine the vertex position. The V0 hodoscopes, placed around the interaction point are used as a trigger, in coincidence with trigger signal provided by muons detected in the muon arm. In addition, V0 are used with the Zero Degree Calorimeter (ZDC) for background rejection. The T0 Cerenkov detectors, also placed around the interaction point are used for luminosity calculation. When measuring J/$\psi$ with the muon arm, one can only measure inclusive J/$\psi$, which include both prompt J/$\psi$ coming from direct production and higher states decay, and non-prompt J/$\psi$ coming mainly from B-meson decay.

\section{Previous ALICE results}
Effects of the QGP on the charmonium production are quantified using the nuclear modification factor $R$\textsubscript{AA}, which is the ratio of the yield of charmonium in Pb-Pb collisions over the yield of charmonium in pp collisions, normalized by the number of binary collisions N\textsubscript{coll}. $R$\textsubscript{AA} can also be defined through the nuclear overlap function T\textsubscript{AA}. If the Pb-Pb collisions are just a superposition of pp collisions, then we would have $R\textsubscript{AA} = 1$. On the contrary, $R\textsubscript{AA} \neq 1$ means that nuclear and/or QGP effects are presents.
In 2011, ALICE published the J/$\psi$ $R$\textsubscript{AA} as a function of centrality at nucleon-nucleon collision energy of $\sqrt s\textsubscript{NN} = 2.76$ TeV \cite{Abelev:2013ila}. The result is shown in Figure 1 compared to the results obtained by PHENIX in Au-Au collisions at $\sqrt s\textsubscript{NN} = 0.2$ TeV. In both case a clear suppression of the J/$\psi$ is observed, however the suppression is smaller for central events in ALICE, which is an indication of regeneration.

\begin{figure}
\begin{center}
\includegraphics[scale=0.34]{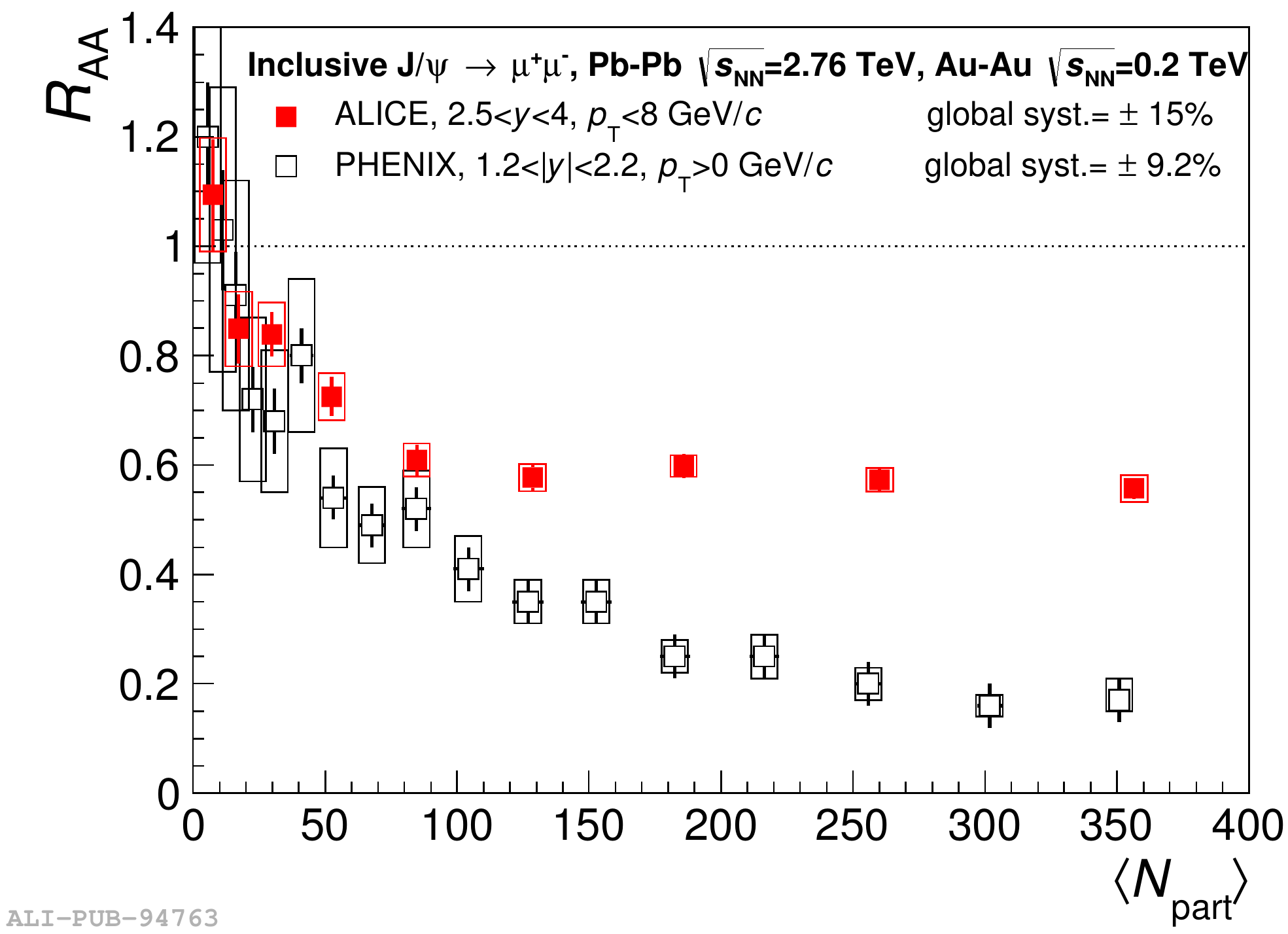}

\caption{$R$\textsubscript{AA} of the J/$\psi$ at $\sqrt s\textsubscript{NN} =2.76$ TeV compared with PHENIX \cite{Abelev:2013ila}}
\end{center}
\label{RAAJPsi2011}
\end{figure}

\section{New Results at $\sqrt s\textsubscript{NN} = 5.02$ TeV}
The data at $\sqrt s\textsubscript{NN} = 5.02$ TeV were taken in December 2015, and correspond to an integrated luminosity $L_{\rm int} = 225$ $\mu$b$\textsuperscript{-1}$. Muon pairs were selected using all the standard cuts described e.g. in \cite{Adam:2016rdg}. Moreover, beam-gas and electromagnetic interactions were rejected using the V0 and ZDC as described previously. The collision's centrality was estimated with a Glauber model fit of the V0 signal amplitude \cite{Adam:2015ptt}.

The J/$\psi$ yield is extracted by fitting the opposite sign dimuon invariant mass spectrum. For the systematic uncertainty on the signal extraction, we test (i) 2 different signal functions, (ii) 2 different methods of dealing with the background, namely a direct empirical fit, or a fit after subtraction of the background using the mixed-event technique, and (iii) several fit ranges. Figure 2 shows examples of the signal extraction for 2 different centrality bins. When integrating over $p\textsubscript{T}$ and considering all centralities, the total number of J/$\psi$ is N\textsubscript{J/$\psi$}$ = 277000$, which is 7 times larger than in \cite{Abelev:2013ila}, due to an increase of luminosity, detector efficiency and J/$\psi$ production cross-section.
 
\begin{figure}
\begin{center}
\includegraphics[scale=0.34]{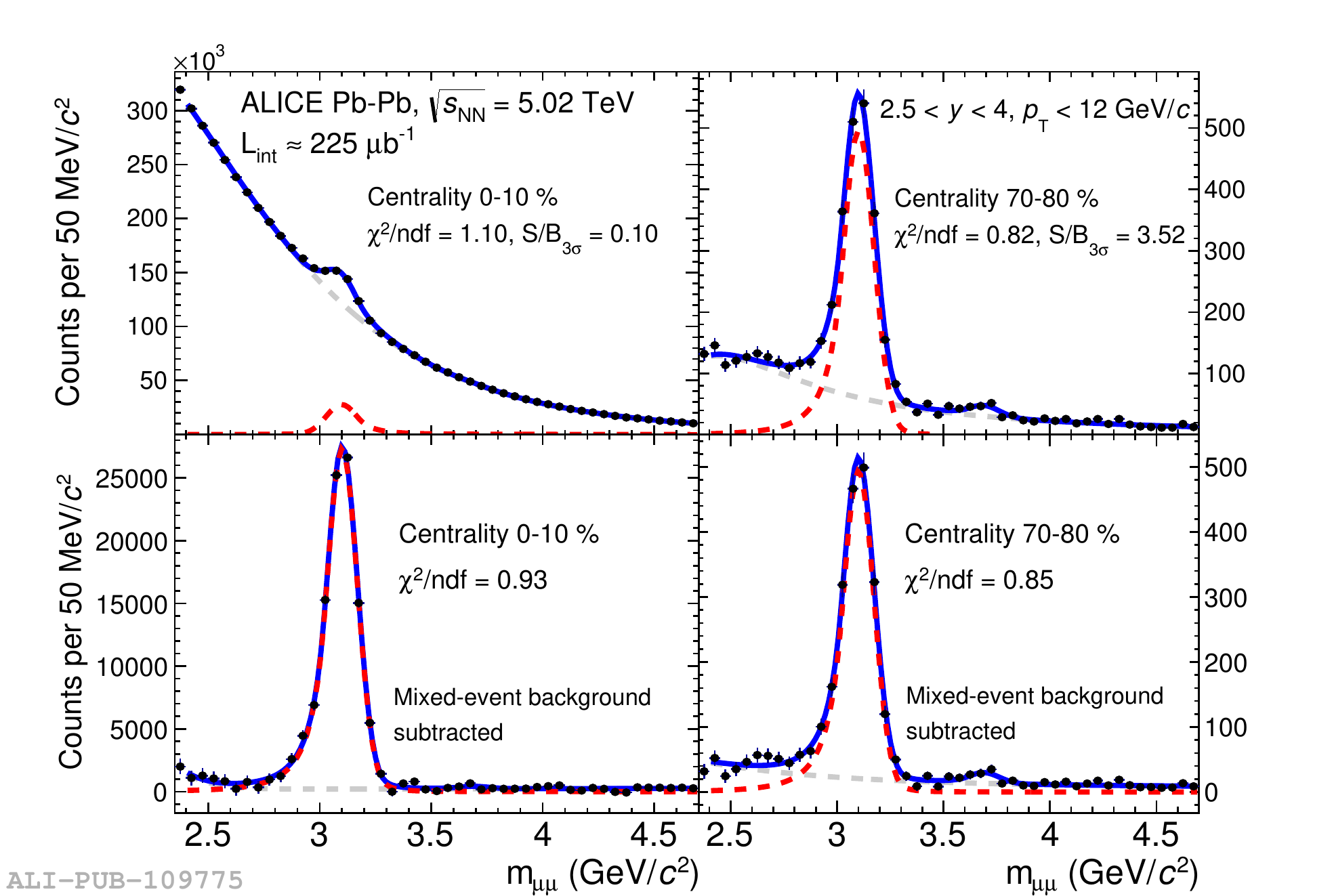}

\caption{Example of signal extraction with a direct empirical fit (up) and the event mixing technique (down) for central events (left) and peripheral events (right) \cite{Adam:2016rdg}}
\end{center}
\label{FitSignal}
\end{figure}

The pp reference, which enters the denominator of $R$\textsubscript{AA}, is measured using the data collected at $\sqrt s = 5$ TeV during 4 days prior to the Pb-Pb collisions, for a total luminosity $L_{\rm int} = 106$ nb$\textsuperscript{-1}$. We obtain an integrated cross section of $\sigma\textsubscript{pp} = 5.61 \pm 0.08(stat) \pm0.28(syst) \mu$b in the transverse momentum interval $0<p\textsubscript{T}<12$ GeV/$c$ and rapidity interval $2.5<y<4.0$.

The acceptance $\times$ efficiency was calculated with Monte-Carlo simulations using the embedding technique to reproduce more accurately the occupancy of the detector.
The nuclear overlap function is calculated using a Glauber model.
Each contribution to the $R$\textsubscript{AA} adds sources of systematic uncertainty - for instance the trigger and tracking efficiency in the acceptance $\times$ efficiency evaluation, the signal extraction, the pp reference. The details of the evaluation can be found in \cite{Adam:2016rdg}. The size of systematic uncertainties amounts to about 8\% for the $R$\textsubscript{AA}.
Figure 3 shows the resulting J/$\psi$ $R$\textsubscript{AA} as a function of centrality, compared with the previous results at $\sqrt s\textsubscript{NN} =2.76$ TeV. The increased statistics allows to use finer bins in centrality, and a more stable behaviour of the detectors to greatly reduce the systematic uncertainties with respect to the result from \cite{Abelev:2013ila}. 

 \begin{figure}
\begin{center}
\includegraphics[scale=0.34]{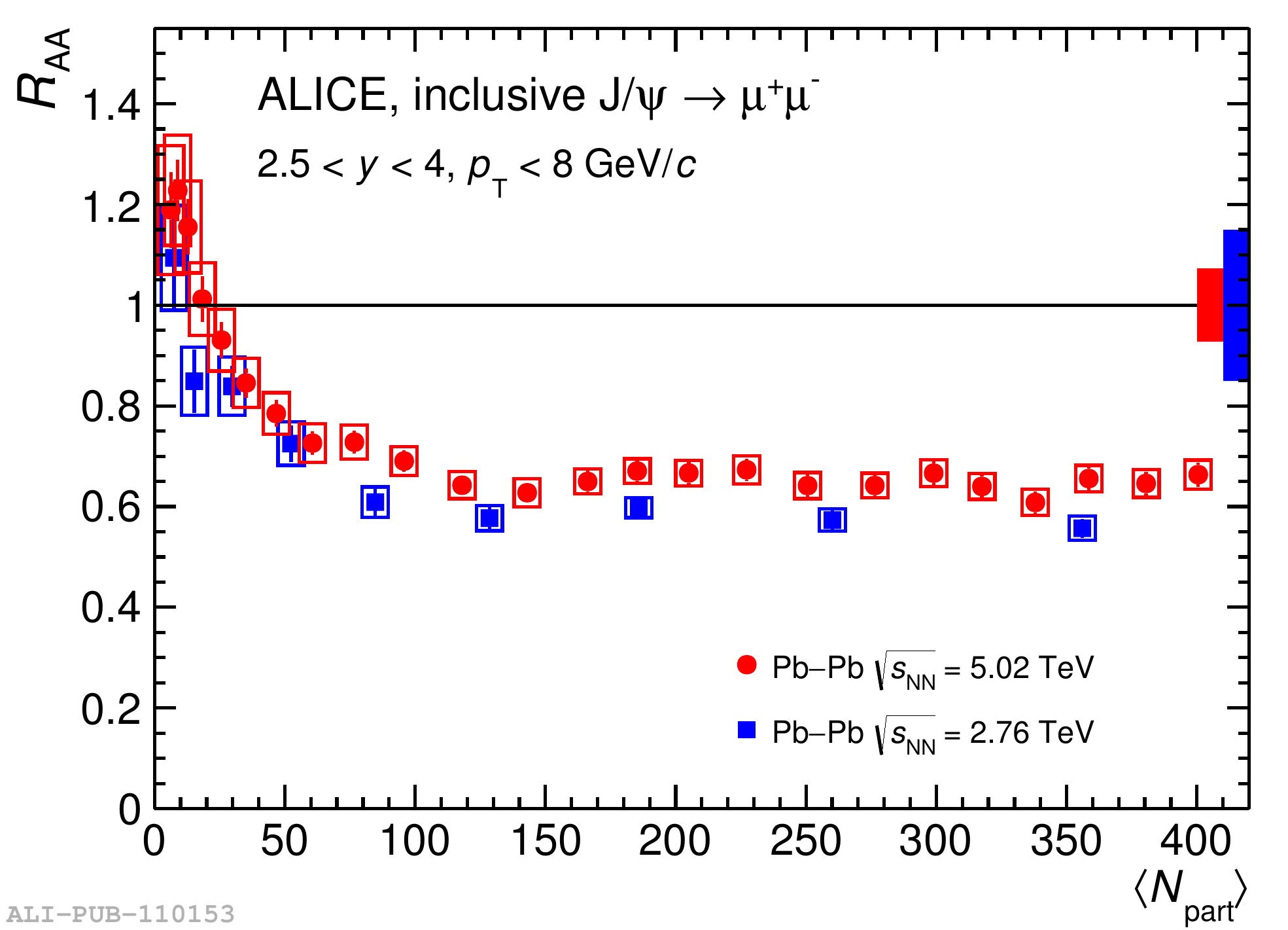}

\caption{$R$\textsubscript{AA} of the J/$\psi$ at $\sqrt s\textsubscript{NN} =5.02$ TeV compared with  $\sqrt s\textsubscript{NN} =2.76$ \cite{Adam:2016rdg}}
\end{center}
\label{RAAManyBins}
\end{figure}
A clear suppression of the J/$\psi$ is observed, with almost no centrality dependence for $N\textsubscript{part}>100$, where $N\textsubscript{part}$ is the number of nucleons taking part to the collision. Two extreme assumptions can be performed to estimate the non-prompt J/$\psi$ contribution to the inclusive $R$\textsubscript{AA}: if the non-prompt J/$\psi$ were completely suppressed ($R\textsubscript{AA}(non$-$prompt)=0$), then the $R$\textsubscript{AA} of the prompt J/$\psi$ would be 10\% higher; if on the contrary the non-prompt J/$\psi$ were not suppressed ($R\textsubscript{AA}(non$-$prompt)=1$), then the $R$\textsubscript{AA} of the prompt J/$\psi$ would be from 5\% to 1\% lower.
Integrating over centrality we have at 5.02 TeV : $R\textsubscript{AA}\textsuperscript{0-90\%}(0< p\textsubscript{T}<8$ GeV/$c) = 0.66 \pm 0.01(stat.) \pm 0.05 (syst.)$
Compared with the 2.76 TeV, we obtain the ratio : $R\textsubscript{AA}\textsuperscript{0-90\%}(5.02$ TeV$)/R\textsubscript{AA}\textsuperscript{0-90\%}(2.76$ TeV$) = 1.13 \pm 0.02(stat.) \pm 0.18 (syst.)$
Results at 2.76 TeV and 5.02 TeV are compatible within uncertainties.

In peripheral collisions, we observe that  $R\textsubscript{AA}>1$, which is explained by an excess of J/$\psi$ produced at very low $p\textsubscript{T}$. This excess is likely caused by the photoproduction of J/$\psi$ in Pb-Pb collisions \cite{Adam:2015gba}. Applying a cut on the J/$\psi$ transverse momentum $p\textsubscript{T}>0.3$ GeV, 80\% of these J/$\psi$ are removed, and the result is then better suited to compare the data to theory models, since they do not include photoproduction. Figure 4 shows the  R\textsubscript{AA}of the J/$\psi$ compared with different models. The first model is the Statistical Hadronization Model (SHM) \cite{0954-3899-38-12-124081}. In this model, the primordial charmonia are completely suppressed in the QGP and the charmonium production occurs at phase boundary by statistical hadronization of the charm quarks. The second model is the Comover Interaction Model (CIM) \cite{Ferreiro201457} : dissociation occurs by interaction with a dense comoving partonic medium, and regeneration is added as a gain term to the comover dissociation. Finally, the data are also compared to two transport models (TM) \cite{ZHAO2011114},\cite{PhysRevC.89.054911}, in which there is a continuous charmonium dissociation-regeneration in the QGP, which is described by a rate equation. 
 
\begin{figure}
\begin{center}
\includegraphics[scale=0.34]{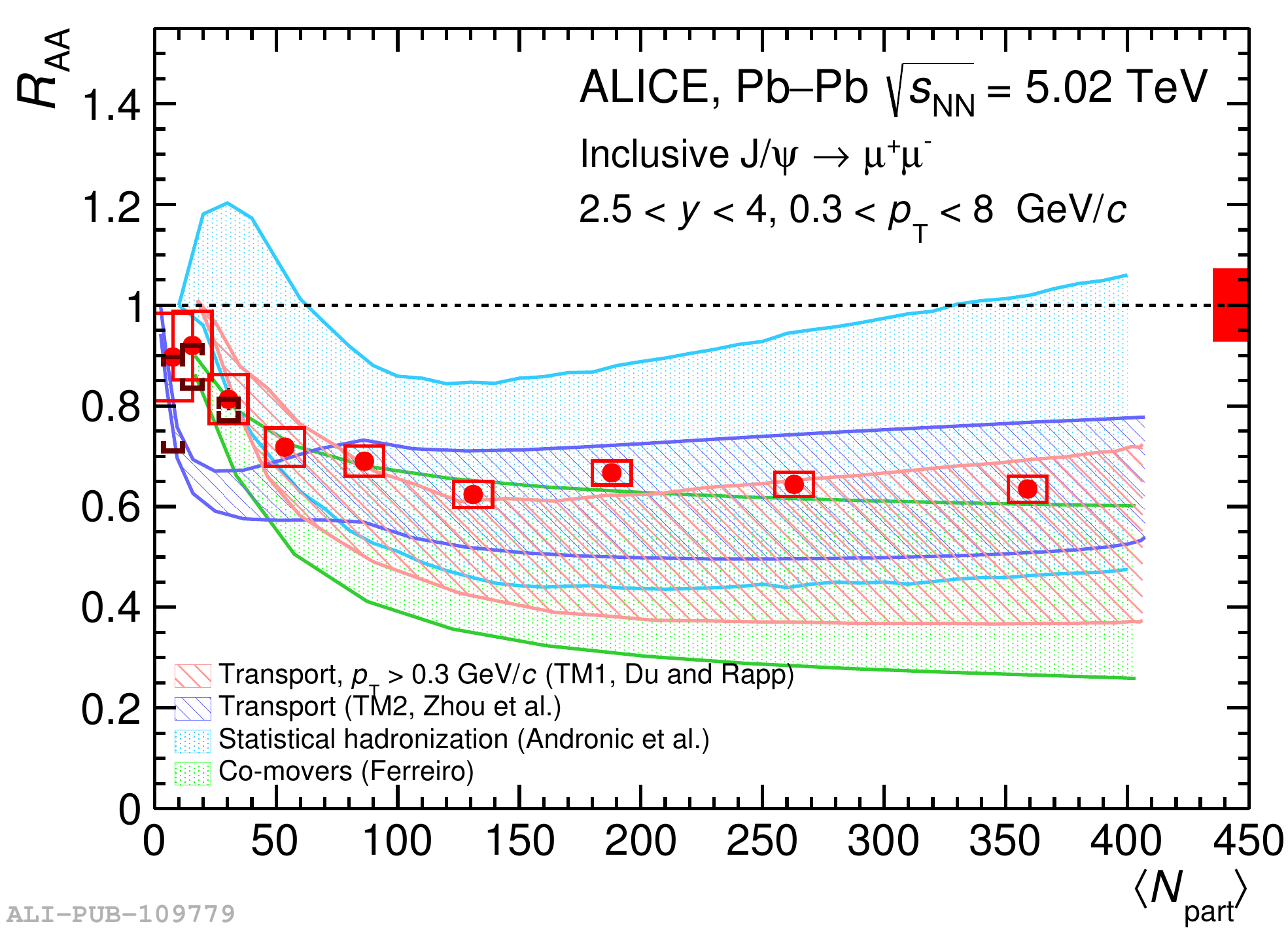}

\caption{$R$\textsubscript{AA} of the J/$\psi$ at $\sqrt s\textsubscript{NN} =5.02$ TeV compared with different theoretical models \cite{Adam:2016rdg}}
\end{center}
\label{RAATheory}
\end{figure}

A large spread among the different calculations, as well as large theoretical uncertainties for each model can be observed. These uncertainties are mainly due to the choice of the $\rm{c\bar{c}}$ cross-section and cold nuclear matter effects. For comover and transport models, a better agreement is found with the upper limits, which corresponds to an absence of nuclear shadowing. Nuclear shadowing is a cold nuclear matter effect observed among other in p-Pb collisions in ALICE, therefore an absence of nuclear shadowing is an extreme assumption.
Each of the models uses a different value for $\sigma\textsubscript{$\rm{c\bar{c}}$}$, and none can be ruled out by the data.

The ratio of the nuclar modification factors between $\sqrt s\textsubscript{NN} =2.76$ and 5.02 TeV (also called double-ratio) can be studied, in order to cancel out some of the uncertainties on the models; in the data only the uncertainty on T\textsubscript{AA} is cancelled out when forming this double-ratio. The results are shown in Figure 5, and the  uncertainties on the models correspond to a 5\% variation on $\sigma_{\rm{c\bar{c}}}$. When removing the non-prompt J/$\psi$ contribution, using the same two assumptions for the non-prompt $R$\textsubscript{AA} as described above, the double ratio varies within 2\%. All three models are compatible with the data, and we observe no significant centrality dependance of the ratio.
 
\begin{figure}
\begin{center}
\includegraphics[scale=0.34]{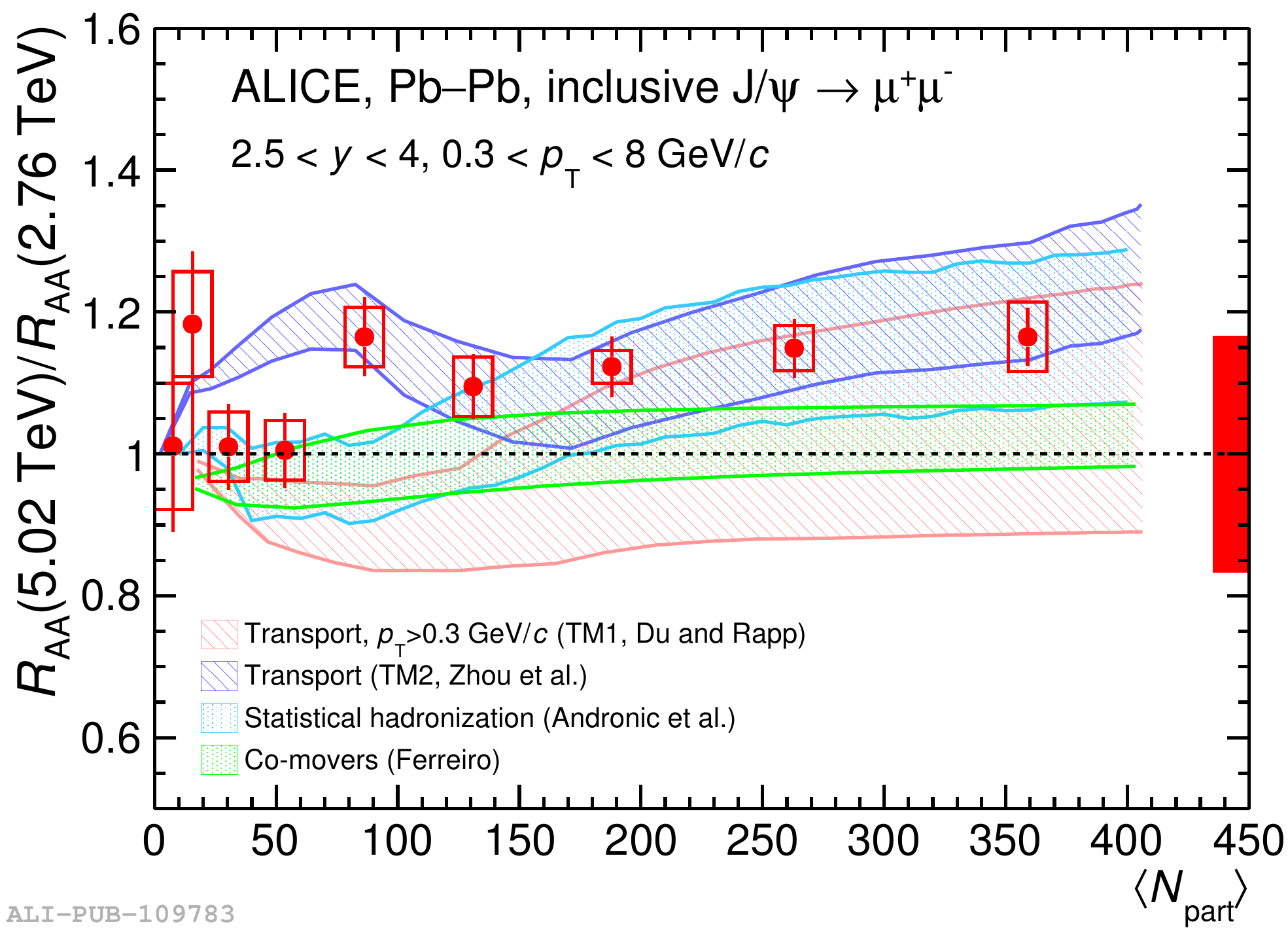}

\caption{Double ratio of the J/$\psi$ compared with theory \cite{Adam:2016rdg}}
\end{center}
\label{rTheory}
\end{figure}

Finally, the $p\textsubscript{T}$ dependance of the $R$\textsubscript{AA} compared to the Transport Model is shown in Figure 6. The $p\textsubscript{T}$ range has been extended to 12 GeV/$c$ with respect to the results from \cite{Abelev:2013ila}. A smaller suppression is observed at low $p\textsubscript{T}$ than at high $p\textsubscript{T}$, as expected from models with a strong regeneration component. A hint of an increase of the R\textsubscript{AA} with respect to $\sqrt s\textsubscript{NN} =2.76$ TeV is also observed between 2 and 6 GeV/$c$. Details on the results can be found in \cite{Adam:2016rdg}.
\begin{figure}
\begin{center}
\includegraphics[scale=0.34]{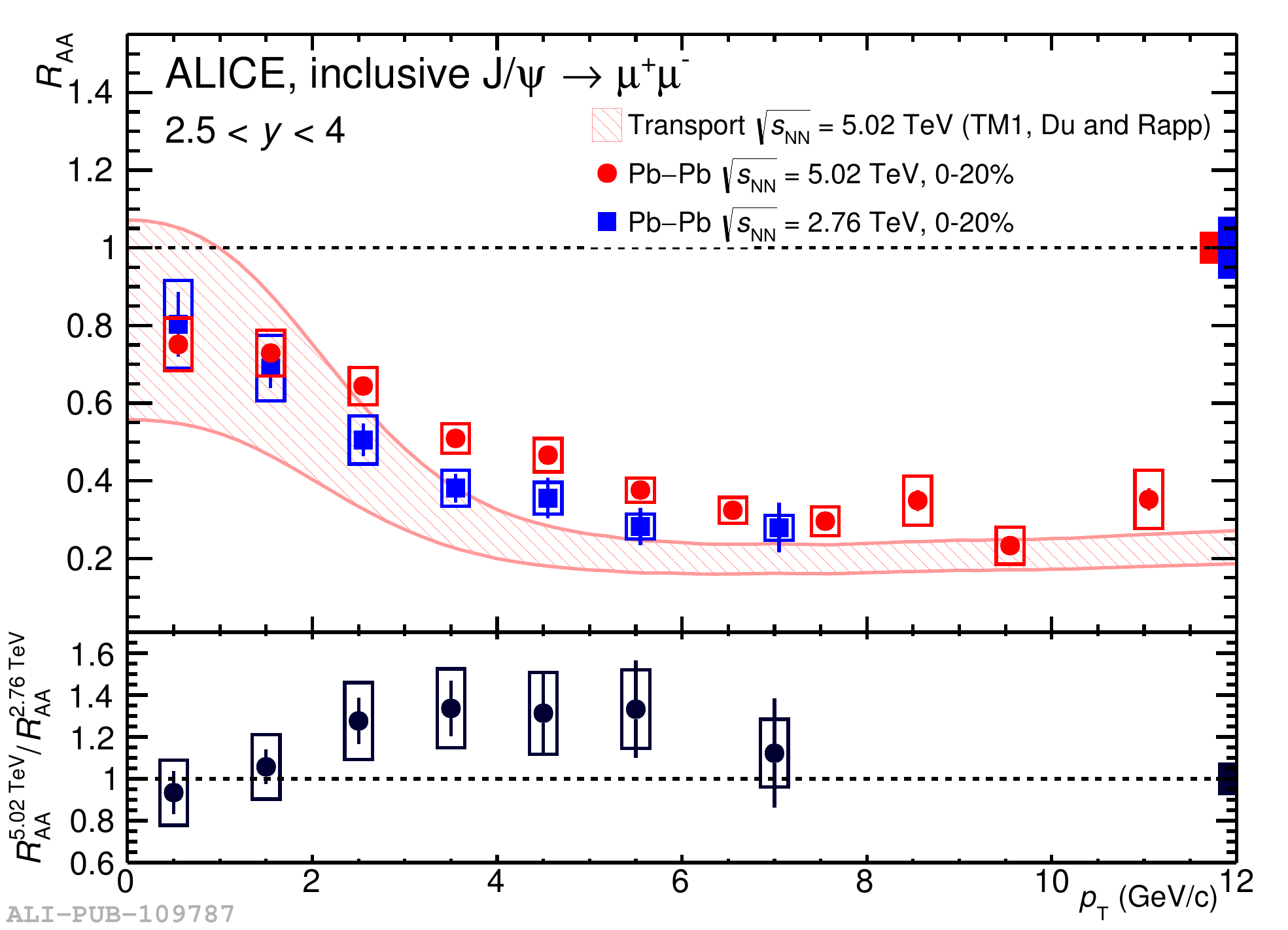}

\caption{$R$\textsubscript{AA} of the J/$\psi$ at $\sqrt s\textsubscript{NN} =5.02$ TeV vs $p\textsubscript{T}$ compared with different theory \cite{Adam:2016rdg}}
\end{center}
\label{rTheory}
\end{figure} 

\section{Conclusion}
The inclusive nuclear modification factor of the J/$\psi$ in Pb-Pb collisions at $\sqrt s\textsubscript{NN} =5.02$ TeV at forward rapidity has been measured down to $p\textsubscript{T} = 0$.
The centrality and p\textsubscript{T} dependence of the $R$\textsubscript{AA} have been studied and show an increase of the J/$\psi$ suppression up to N\textsubscript{part} = 100 followed by a plateau, and an increase of the J/$\psi$ suppression at high p\textsubscript{T} with respect to low p\textsubscript{T}. The comparison with the results at $\sqrt s\textsubscript{NN} =2.76$ TeV shows that results are compatible within uncertainties in the full centrality range, and we observe a hint of an increase with colliding energy for $R$\textsubscript{AA} as a function of $p\textsubscript{T}$ for $2<p\textsubscript{T}<6$ GeV/$c$. These results are compatible within uncertainties with theoretical models and support a picture of J/$\psi$ suppression and regeneration competing in the QGP.




\nocite{*}
\bibliographystyle{elsarticle-num}
\bibliography{Feuillard_V.bib}







\end{document}